\documentclass{article}
\usepackage{spconf}
\usepackage{amsmath,graphicx}
\usepackage{hyperref}
\usepackage{amsfonts}

\newcommand{\norm}[1]{\left\lVert#1\right\rVert}

\title{The PCG-AIID System for L3DAS22 Challenge: \\
MIMO and MISO convolutional recurrent Network \\ 
for Multi Channel Speech Enhancement and Speech Recognition}
\name{Jingdong Li$^*$, Yuanyuan Zhu$^*$, Dawei Luo$^*$, Yun Liu, Guohui Cui, Zhaoxia Li}
\address{AI Interaction Division, Tencent PCG \\ 
\{jingdongli, solverzhu, daweilluo\}@tencent.com}
\begin{document}
\maketitle
%
\def\thefootnote{*}\footnotetext{These authors contributed equally to this work.}\def\thefootnote{\arabic{footnote}}
\begin{abstract}
This paper described the PCG-AIID system for L3DAS22 challenge in Task 1: 3D speech enhancement in office reverberant environment. We proposed a two-stage framework to address multi-channel speech denoising and dereverberation. In the first stage, a multiple input and multiple output (MIMO) network is applied to remove background noise while maintaining the spatial characteristics of multi-channel signals. In the second stage, a multiple input and single output (MISO) network is applied to enhance the speech from desired direction and post-filtering. As a result, our system ranked 3rd place in ICASSP2022 L3DAS22 challenge and significantly outperforms the baseline system, while achieving 3.2\% WER and 0.972 STOI on the blind test-set.
\end{abstract}
\section{Introduction}
\label{sec:intro}
Multi-channel speech enhancement methods have attracted a lot of attention over the past decades. Microphone arrays consisting of multiple microphones can capture much richer information of the target speech signal, compared to single microphone. It is more suitable for solving speech enhancement problems in complex acoustic scenarios using multi-channel processing technique.

Conventionally, the main stream of multi-channel speech enhancement is acoustic beamforming, which has different sensitivities to signals from different directions of arrival. Therefore, it can enhance signals from target direction while attenuating signals from other directions. Given a target direction, delay-and-sum beamformer \cite{vanbf} shifts each microphone signal, and then sums these time-shifted signals to produce the output. Another widely used adaptive beamformer is the minimum variance distortion-free response (MVDR) beamformer \cite{capon}, which minimizes the average energy of the beamformer output while preserving the undistorted signal from the target direction. 

Recently, DNN-based time-frequency (T-F) masking or mapping has been established as the mainstream approach for single-channel speech enhancement and separation \cite{overview,mapping}. To address the multi-channel speech separation problem, a single channel noise suppression network is used in \cite{maskbf1,maskbf2,maskbf3} to generate time-frequency masks. The estimated mask is then used to compute the spatial convariance matrix and steer vector. Compared to traditional algorithms, these methods provides a more accurate estimate of the beamforming weights. However, these methods only estimate a magnitude-domain time-frequency mask using single-channel speech. The inter-channel features are not fully exploited.

\begin{figure*}[ht]
\centering
\includegraphics[width=\textwidth]{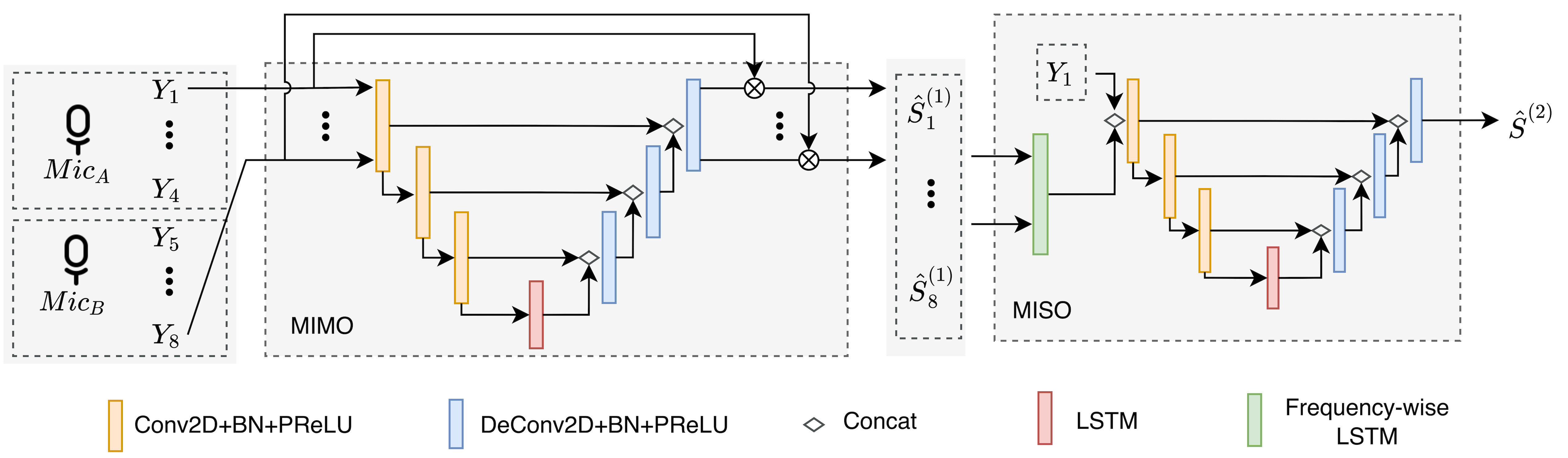}
\caption{System diagram.}
\label{fig:f1}
\end{figure*}

For microphone arrays with fixed geometry, the spatial and time-frequency features can be simultaneously utilized by the neural network. ADL-MVDR\cite{adl} used a recurrent neural network (RNN) to directly estimate the frame-level inverse of the convariance and steer vector, resulting in less residual noise and higher ASR accuracy. The multi channel complex spectral mapping \cite{wzq_mch_chime,wzq_mch_reverb} has also shown effectiveness in speech enhancement and separation. The complex spectrograms of the multi-channel mixture are concatenated and fed into a DNN to directly estimate target speech, while the spatial cues are implicitly utilized by the neural network. Similarly, the baseline system of the ICASSP 2022 L3DAS22 Challenge\cite{l3das}, MMUB \cite{renmimo} also adopts a multi channel neural beamforming. Specifically, the concatenation of multi channel mixture is fed in to a UNet to estimate a complex weight for each channel. Then the sum of the weighted mixtures is the estimated target signal. It also does not require explicit spatial feature such as the interaural phase difference(IPD). 

In this paper, we proposed two-stage framework to address multi channel speech enhancement problem for microphone array with fixed geometry. The first stage performs coarse time-frequency filtering, which estimates the multi-channel target speech. The estimated multi-channel speech still implicitly maintains the spatial cues. Then the second stage performs the spatial filtering and further post-filtering. The proposed method significantly outperformed the baseline MMUB in terms of STOI and WER, and ranked the 3rd place of the L3DAS22 challenge.

The rest of the paper is organized as follows. In section \ref{sec:methods}, we introduce the overall diagram and details of the proposed methods. In section \ref{sec:exp}, we compare the proposed system with other models. Conclusions are drawn in section \ref{sec:conclusion}

\section{methods}
\label{sec:methods}

\subsection{Problem formulation}
\label{ssec:problem formulation}
Given the noisy and reverberant $P$ channel signals $\{ y_{p} \}_{p=1}^{P}$, the physical model can be formulated as:
\begin{equation} \label{eq:time}
 y_p = s * h_p + n_p
\end{equation}
where $s$ is the source signal, $h_p$ is the room impulse response and $n_p$ is the background noise, $p$ is the channel index.

With the short-time Fourier transform (STFT), the Eq.(\ref{eq:time}) can be represented as:
\begin{equation} \label{eq:freq}
Y_p(t,f) = S(t, f)H_p(f) + N_p(t, f)
\end{equation}
where $Y(t, f)$, $S(t, f)$, and $N_p(t, f)$ respectively represent the STFT of the mixture, clean source and noise signals.

In this work, we propose a model to recover source signal $S(t, f)$ from multi channel mixture $\{Y_p(t, f)\}_{p=1}^{P}$, which is illustrated in Figure.\ref{fig:f1}. We will introduce each part of the model as follows.


\subsection{Convolutional Recurrent Network}
\label{ssec: crn}
As illustrated in Figure.\ref{fig:f1}, we use convolutional recurrent network (CRN) \cite{gcrn} as the basic structure to estimate single channel and multi-channel complex spectrogram of desired signal. CRN consist of a convolutional encoder-decoder (CED) and a bidirectional long short term memory network (LSTM). The encoder in the CED extracts high-level frequency features, and the decoder reconstruct the frame by gradually upsample. The concatenation of complex spectrogram of the multi-channel signals are fed in to encoder. Then, two separate decoders estimate the real and imaginary spectrogram of target signal.  Each encoder block is consist of a 2d convolution, 2d batch normalization and PReLU activation. And each decoder block is consist of a 2d deconvolution, 2d batch normalization, and PReLU activation.  To track the temporal dynamics of speech, LSTM layers are inserted between the encoder and the decoder. In this work, we used 32ms frame size, 4ms frame shift and 512 point fast Fourier transform (FFT). And the hyper-parameters of the CRN showed in the Table.\ref{tab:crn}, where the $C_\text{out}$  and $C_\text{out}$ denotes the input channel and output channel.

\subsection{Two-stage processing}
The proposed methods consist of two processing stage. The first stage is a MIMO-CRN to remove background noise while maintaining the spatial relation among channels. The second stage is a MISO-CRN. The multi channel estimation $\{\hat{S}^{(1)}_{p}\}_{p=1}^{P}$ from the first stage are fed to the MISO-CRN to generate the final estimation of the target source signal $\hat{S}^{(2)}$.
\subsubsection{Stage I: MIMO-CRN}
\label{sssec:mimo}
For the MIMO-CRN in the first stage, the input channel $C_\text{in}$ and the output channel $C_\text{out}$ are both set to $2P$. The real and imaginary spectrogram of $P$ channel mixture are concatenated as the input feature. And the two separate decoder predict the real and imaginary part of $P$ channel complex ideal ratio mask (cIRM), which will be element-wise multiplied to the STFT of mixture signals to generate the estimation of clean speech $\{\hat{S}^{(1)}_{p}\}_{p=1}^{P}$. Due to the background noised is reduced in the first stage, the further spatial filtering in the second stage is easier to find the direction of the source.

\subsubsection{Stage II: MISO-CRN}
\label{sssec:miso}
For the acoustic beamformers in STFT domain, spatial filtering is independent for each frequency band. According to this prior, a frequency-wise LSTM is adopted to independently perform spatial filtering, which is similar as \cite{igcrn}. We denote the complex spectrogram in band $f$ as  $\mathbf{\hat{S}}(f)^{(1)} \in \mathbb{C}^{T \times P}$, which is produced from the first stage. Then the process of spatial filtering in band $f$ can be formulated as :
\begin{equation}
    \mathbf{X} = \text{Concat}(\{Re(\mathbf{\hat{S}}(f)^{(1)}), Im(\mathbf{\hat{S}}(f)^{(1)})\}) \in \mathbb{R}^{T \times 2P}
\end{equation}
\begin{equation}
    \mathbf{\tilde{X}} = \text{FC}(\text{LSTM}(\text{LN}(\mathbf{X}))) \in \mathbb{R}^{T \times 2}
\end{equation}
where $LN(\cdot)$ and $FC(\cdot)$ respectively denotes the layer normalization and fully connected layer. Then we split the $\mathbf{\tilde{X}}$ to two part, regarded as the real and imaginary part of the complex spectrogram of spatial filtered signal. In this work, we use two layers of bidirectional LSTM with 64 hidden size.

After that, we concatenate the spatial filtered signal and the first channel of mixture signals, then fed it to the second CRN. The second CRN directly predicts the real and the imaginary spectrogram of the target source signal. The input channel $C_\text{in}$ and the output channel $C_\text{out}$ of the second CRN are set to 4 and 2, respectively.

\begin{table}[]
\setlength\tabcolsep{1.5pt} 
\begin{tabular}{|c|c|c|c|}
\hline
layer  & input\_size   & hyper-parameters    & output\_size  \\ \hline
           & ($\bf{C_{in}}$, T, 256) & (1, 3), (1, 2), 16   & (16, T, 128)   \\ \cline{2-4} 
           & (16, T, 128)  & (1, 3), (1, 2), 32   & (32, T, 64)    \\ \cline{2-4} 
Conv2d     & (32, T, 64)   & (1, 3), (1, 2), 64   & (64, T, 32)    \\ \cline{2-4} 
           & (64, T, 32)   & (1, 3), (1, 2), 128  & (128, T, 16)   \\ \cline{2-4} 
           & (128, T, 16)  & (1, 3), (1, 2), 256  & (256, T, 8)    \\ \cline{2-4} 
           & (256, T, 8)   & (1, 3), (1, 2), 256  & (256, T, 4)    \\ \hline
Reshape    & (256, T,  4)  & -                    & (T, 1024)      \\ \hline
LSTM       & (T, 1024)     & 2 layer, 1024        & (T, 1024)      \\ \hline
Reshape    & (T, 1024)     & -                    & (256, T,  4)   \\ \hline
           & (512, T, 4)   & (1, 3), (1, 2), 256  & (256, T, 8)    \\ \cline{2-4} 
           & (512, T, 8)   & (1, 3), (1, 2), 128  & (128, T, 16)   \\ \cline{2-4} 
Deconv2d   & (256, T, 16)  & (1, 3), (1, 2), 64   & (64, T, 32)    \\ \cline{2-4} 
           & (128, T, 32)  & (1, 3), (1, 2), 32   & (32, T, 64)    \\ \cline{2-4} 
           & (64, T, 64)   & (1, 3), (1, 2), 16   & (16, T, 128)   \\ \cline{2-4} 
           & (16, T, 128)  & (1, 3), (1, 2), $\bf{C_{out}}$ & ($\bf{C_{out}}$, T, 256) \\ \hline
\end{tabular}
\caption{Hyper-parameters of the CRN.}
\label{tab:crn}
\end{table}

\subsection{Loss function}
\label{ssec:loss}
We use a combination of the RI-Mag loss and Mag-hurts loss as the optimization target:
\begin{equation}
    \label{eq:loss}
    L = L_{\text{RI-Mag}} + \alpha L_{\text{Mag-hurts}} 
\end{equation}
where the $\alpha$ is a scalar weight. In this work, $\alpha$ is set 2. Recently, the power-compressed spectrum has been proposed in \cite{mag_compress} to improve the performance of dereverberation model. Therefore, we conduct square root compression on magnitude of estimated speech $\hat{S}$ and target speech $S$. For convenience, we still represent the compressed estimated speech and target speech using $\hat{S}$ and $S$. 

The RI-Mag loss calculated by comparing the real, imaginary and magnitude spectrograms between  $\hat{S}$ and $S$:
\begin{equation}
    L_{\text{RI}} = \norm{Re(S) - Re(\hat{S})}^2 + \norm{Im(S) - Im(\hat{S})}^2
\end{equation}
\begin{equation}
    L_{\text{Mag}} = \norm{Mag({S}) - Mag(\hat{S})}^2
\end{equation}
\begin{equation}
    \label{eq:rimag}
    L_{\text{RI-Mag}} = L_{\text{RI}} + L_{\text{Mag}}
\end{equation}
where $\norm{ \cdot }^2$ denotes the $l2$ norm, $Re(\cdot)$, $Im(\cdot)$ and $Mag(\cdot)$ respectively denotes real part, imaginary part and magnitude of a complex number.

As the magnitude is more important than phase in speech, the Mag-hurts loss is adopted to further penal the over-suppression of target magnitude spectrogram.
\begin{equation}
    \label{eq:maghurts}
    L_{\text{Mag-hurts}} =  \norm{Max(0, Mag({S}) - Mag(\hat{S}))}^2
\end{equation}

\section{Experiments and Results}
\label{sec:exp}

\subsection{Dataset}
\label{ssec:dataset}
The proposed system evaluated on the dataset provided by the L3DAS22 challenge. The multi-channel signals are simulated by convolving the monaural clean speech and multi-channel impulse response (IR), then add the background noise convolved with another different IR.

The clean speech comes from librispeech corpus \cite{librispeech}, and noise signals comes from the FSD50K \cite{fsd50k}.

The IRs are captured in a large office room with dimensions 6m (length) by 5m (width) by 3m (height). Two ambisonics microphone array (MicA, MicB) are placed in the center of the room. Each microphone array has 4 microphones, and produces 4 channel signals. The distance between micA and MicB is 20cm, which is close to the distance between human ears. Both microphone array positioned at the same height 1.3m, which is close to the ear height of a seated person. The position of source signal is randomly selected from different 252 positions. 

\subsection{Evaluation metrics}
\label{ssec:metrics}
The evaluation metric for this task is a combination of the short time objective intelligibility (STOI) and word error rate (WER). The WER are computed by the Wav2Vec \cite{wav2vec} model, which pretrained on 960h Librispeech corpus. The final metric is defined as:
\begin{equation} \label{eq:metric}
    \text{metric} = (\text{STOI} + (1 - \text{WER})) / 2
\end{equation}
The STOI lies in the 0-1 range and higher values are better. And the WER lies in the 0-1 range and lower values are better. Therefore, the final metric lies in the 0-1 range and higher values are better.

\subsection{Training setup}
We trained models with the batch-size of 64 on NVIDIA A100 GPUs, and use AdamW gradient optimizer to update the parameters. The initial learning rate is 0.001, then learing-rate reduced by 0.5 every 50,000 iterations. The weight decay of AdamW is set to 0.01. 

\subsection{Baseline systems}
We compared the proposed system with the MMUB, which won the 1st prize of Task1 in the L3DAS21 challenge. The MMUB \cite{renmimo} use the multi-channel UNet to estimate beamforming filters, which is then element-wise multiplied with the multi-channel complex spectrogram. Then, the estimation of the target signal is sum over the filtered multi-channel signals. 

We also compared the proposed method with a combination algorithm of weight prediction error (WPE) and mask MVDR. WPE is one of the most popular dereverberation algorithm. It estimates the late reverberation as a linear combination of the past observations, then subtracts the reverberation from the mixture. MVDR is a widely used adaptive beamformer for multi-channel speech enhancement. MVDR aims to minimize the output power with the constraint that the gain of signal arriving from the target direction is 1. A NN-based time-frequency mask estimator is utilized to provided a more accurate convariance matrix of speech and noise for MVDR. 

\subsection{Results}
\label{ssec:results}

We firstly compared the proposed methods with baseline methods on the development set of the L3DAS22 challenge. The results comparison in terms of STOI and WER among different methods are shown in Table.\ref{tab:dev}. 
The combination of the WPE and MVDR gets a minor improvement of STOI and WER over the noisy mixture. Because the signal estimated by the MVDR beamformer is suffered from residual noise, which is harmful for both intelligibility and speech recognition. To show the advantage of multi-channel processing, we compare single-channel denoising CRN and multi-channel denoising networks. It is obvious that the multi-channel methods MISO-CRN-4ch and MISO-CRN-8ch yield the better STOI and WER than the single-channel model CRN-1ch. And MISO-CRN-4ch and MISO-CRN-8ch has comparable performance, which indicates performance may be saturated as the increasing of number of micphones. We also investigated that using CRN to estimate the beamforming weights like MMUB. By replacing the UNet with CRN, the MISO-CRN-8ch-filtersum methods yield 0.046 STOI and 0.071 WER improvements over the baseline MMUB. The proposed two-stage framework has the best performance on the development dataset. The MIMO-MISO-CRN yileds 0.067 STOI and 0.108 WER improvements over MMUB, showing the superiority of our models.

In the evaluation of the L3DAS22 challenge, we achieved 0.972 STOI and 0.032 WER, which ranked the 3rd place. The Table.\ref{tab:test} showed the metrics of top-5 models and baseline system on the test-set of L3DAS22.

\begin{table}[]
\begin{tabular}{lccc}
\hline
\textbf{Model}           & \multicolumn{1}{l}{\textbf{STOI}} & \multicolumn{1}{l}{\textbf{WER}} & \multicolumn{1}{l}{\textbf{Metric}} \\ \hline
Mixture (ch0)            & 0.615                             & 0.389                            & 0.613                               \\
MMUB (baseline)          & 0.908                             & 0.144                            & 0.882                               \\ \hline
WPE + MVDR(frame)        & \multicolumn{1}{l}{0.67}          & \multicolumn{1}{l}{0.313}        & \multicolumn{1}{l}{0.678}           \\
WPE + MVDR(block)        & \multicolumn{1}{l}{0.641}         & \multicolumn{1}{l}{0.333}        & \multicolumn{1}{l}{0.654}           \\
CRN-1ch                  & 0.869                             & 0.215                            & 0.827                               \\
MISO-CRN-4ch             & 0.917                             & 0.140                            & 0.888                               \\
MISO-CRN-8ch             & 0.912                             & 0.150                            & 0.881                               \\
MISO-CRN-8ch-filtersum   & 0.954                             & 0.073                            & 0.940                               \\
MIMO-MISO-CRN (proposed) & \textbf{0.975}                    & \textbf{0.036}                   & \textbf{0.970}                      \\ \hline
\end{tabular}
\caption{The metrics of different systems on the dev-set}
\label{tab:dev}
\end{table}

\begin{table}[]
\begin{tabular}{clccc}
\hline
\multicolumn{1}{l}{\textbf{Rank}} & \textbf{Model}               & \multicolumn{1}{l}{\textbf{STOI}} & \multicolumn{1}{l}{\textbf{WER}} & \multicolumn{1}{l}{\textbf{Metric}} \\ \hline
1                                 & ESP-SE                       & \multicolumn{1}{l}{0.987}         & \multicolumn{1}{l}{0.019}        & \multicolumn{1}{l}{0.984}           \\
2                                 & BaiduSpeech                  & \multicolumn{1}{l}{0.975}         & \multicolumn{1}{l}{0.025}        & \multicolumn{1}{l}{0.975}           \\
3                                 & \textbf{PCG-AIID (proposed)} & 0.972                             & 0.032                            & 0.970                               \\
4                                 & NPU                          & 0.955                             & 0.047                            & 0.954                               \\
5                                 & BBD                          & 0.945                             & 0.049                            & 0.948                               \\
-                                 & MMUB (baseline)              & 0.878                             & 0.212                            & 0.833                               \\ \hline
\end{tabular}
\caption{The metrics of top-5 systems and baseline on the blind test-set}
\label{tab:test}
\end{table}

\section{Conclusion}
\label{sec:conclusion}
In this paper, a two stage framework is proposed for multi-channel speech enhancement and recognition. Experiment results show the the proposed methods has promising performance for both speech intelligibility and speech recognition. The proposed method has ranked 3rd place of the ICASSP L3DAS22 challenge, significantly outperforms the baseline system MMUB in terms of STOI and WER.
   
Although all top-ranked models in L3DAS22 achieved quite high STOI and very low WER. But we think it is still some distance from practical application. For example, the IRs used for training and testing are sampled from a same office room. We don't know if the model will work well in another unseen room. The robustness of the neural based beamformers still needs to be studied.

\bibliographystyle{IEEEbib}
\bibliography{refs}

\begin{thebibliography}{10}

\bibitem{vanbf}
Barry~D Van~Veen and Kevin~M Buckley,
\newblock ``Beamforming: A versatile approach to spatial filtering,''
\newblock {\em IEEE assp magazine}, vol. 5, no. 2, pp. 4--24, 1988.

\bibitem{capon}
Jack Capon,
\newblock ``High-resolution frequency-wavenumber spectrum analysis,''
\newblock {\em Proceedings of the IEEE}, vol. 57, no. 8, pp. 1408--1418, 1969.

\bibitem{overview}
DeLiang Wang and Jitong Chen,
\newblock ``Supervised speech separation based on deep learning: An overview,''
\newblock {\em IEEE/ACM Transactions on Audio, Speech, and Language
  Processing}, vol. 26, no. 10, pp. 1702--1726, 2018.

\bibitem{mapping}
Yong Xu, Jun Du, Li-Rong Dai, and Chin-Hui Lee,
\newblock ``An experimental study on speech enhancement based on deep neural
  networks,''
\newblock {\em IEEE Signal processing letters}, vol. 21, no. 1, pp. 65--68,
  2013.

\bibitem{maskbf1}
Jahn Heymann, Lukas Drude, and Reinhold Haeb-Umbach,
\newblock ``Neural network based spectral mask estimation for acoustic
  beamforming,''
\newblock in {\em 2016 IEEE International Conference on Acoustics, Speech and
  Signal Processing (ICASSP)}. IEEE, 2016, pp. 196--200.

\bibitem{maskbf2}
Takuya Higuchi, Nobutaka Ito, Takuya Yoshioka, and Tomohiro Nakatani,
\newblock ``Robust mvdr beamforming using time-frequency masks for
  online/offline asr in noise,''
\newblock in {\em 2016 IEEE International Conference on Acoustics, Speech and
  Signal Processing (ICASSP)}. IEEE, 2016, pp. 5210--5214.

\bibitem{maskbf3}
Hakan Erdogan, John~R Hershey, Shinji Watanabe, Michael~I Mandel, and Jonathan
  Le~Roux,
\newblock ``Improved mvdr beamforming using single-channel mask prediction
  networks.,''
\newblock in {\em Interspeech}, 2016, pp. 1981--1985.

\bibitem{adl}
Zhuohuang Zhang, Yong Xu, Meng Yu, Shi-Xiong Zhang, Lianwu Chen, and Dong Yu,
\newblock ``Adl-mvdr: All deep learning mvdr beamformer for target speech
  separation,''
\newblock in {\em ICASSP 2021-2021 IEEE International Conference on Acoustics,
  Speech and Signal Processing (ICASSP)}. IEEE, 2021, pp. 6089--6093.

\bibitem{wzq_mch_chime}
Zhong-Qiu Wang, Peidong Wang, and DeLiang Wang,
\newblock ``Complex spectral mapping for single-and multi-channel speech
  enhancement and robust asr,''
\newblock {\em IEEE/ACM transactions on audio, speech, and language
  processing}, vol. 28, pp. 1778--1787, 2020.

\bibitem{wzq_mch_reverb}
Zhong-Qiu Wang and DeLiang Wang,
\newblock ``Multi-microphone complex spectral mapping for speech
  dereverberation,''
\newblock in {\em ICASSP 2020-2020 IEEE International Conference on Acoustics,
  Speech and Signal Processing (ICASSP)}. IEEE, 2020, pp. 486--490.

\bibitem{l3das}
Eric Guizzo, Christian Marinoni, Marco Pennese, Xinlei Ren, Xiguang Zheng, Chen
  Zhang, Bruno Masiero, Aurelio Uncini, and Danilo Comminiello,
\newblock ``L3das22 challenge: Learning 3d audio sources in a real office
  environment,''
\newblock in {\em 2022 IEEE International Conference on Acoustics, Speech and
  Signal Processing (ICASSP)}. IEEE, 2022.

\bibitem{renmimo}
Xinlei Ren, Lianwu Chen, Xiguang Zheng, Chenglin Xu, Xu~Zhang, Chen Zhang,
  Liang Guo, and Bing Yu,
\newblock ``A neural beamforming network for b-format 3d speech enhancement and
  recognition,''
\newblock in {\em 2021 IEEE 31st International Workshop on Machine Learning for
  Signal Processing (MLSP)}. IEEE, 2021, pp. 1--6.

\bibitem{gcrn}
Ke~Tan and DeLiang Wang,
\newblock ``Learning complex spectral mapping with gated convolutional
  recurrent networks for monaural speech enhancement,''
\newblock {\em IEEE/ACM Transactions on Audio, Speech, and Language
  Processing}, vol. 28, pp. 380--390, 2019.

\bibitem{igcrn}
Jinjiang Liu and Xueliang Zhang,
\newblock ``{Inplace Gated Convolutional Recurrent Neural Network for
  Dual-Channel Speech Enhancement},''
\newblock in {\em Proc. Interspeech 2021}, 2021, pp. 1852--1856.

\bibitem{mag_compress}
Andong Li, Chengshi Zheng, Renhua Peng, and Xiaodong Li,
\newblock ``On the importance of power compression and phase estimation in
  monaural speech dereverberation,''
\newblock {\em JASA Express Letters}, vol. 1, no. 1, pp. 014802, 2021.

\bibitem{librispeech}
Vassil Panayotov, Guoguo Chen, Daniel Povey, and Sanjeev Khudanpur,
\newblock ``Librispeech: an asr corpus based on public domain audio books,''
\newblock in {\em 2015 IEEE international conference on acoustics, speech and
  signal processing (ICASSP)}. IEEE, 2015, pp. 5206--5210.

\bibitem{fsd50k}
Eduardo Fonseca, Xavier Favory, Jordi Pons, Frederic Font, and Xavier Serra,
\newblock ``Fsd50k: an open dataset of human-labeled sound events,''
\newblock {\em arXiv preprint arXiv:2010.00475}, 2020.

\bibitem{wav2vec}
Alexei Baevski, Yuhao Zhou, Abdelrahman Mohamed, and Michael Auli,
\newblock ``wav2vec 2.0: A framework for self-supervised learning of speech
  representations,''
\newblock {\em Advances in Neural Information Processing Systems}, vol. 33, pp.
  12449--12460, 2020.

\end{thebibliography}

\end{document}